\documentclass[12pt,preprint]{emulateapj}
\usepackage{graphicx}

\shorttitle{TERMS Photometry of Known Transiting Exoplanets}
\shortauthors{Dragomir et al.}
\slugcomment{Submitted for publication in the Astronomical Journal}

\begin{document}

\title{TERMS Photometry of Known Transiting Exoplanets}

\author{Diana Dragomir\altaffilmark{1,2},
  Stephen R. Kane\altaffilmark{1},
  Genady Pilyavsky\altaffilmark{3},
  Suvrath Mahadevan\altaffilmark{3,4},
  David R. Ciardi\altaffilmark{1},
  J. Zachary Gazak\altaffilmark{6}
  Dawn M. Gelino\altaffilmark{1},
  Alan Payne\altaffilmark{1},
  Markus Rabus\altaffilmark{5},
  Solange V. Ramirez\altaffilmark{1},
  Kaspar von Braun\altaffilmark{1},
  Jason T. Wright\altaffilmark{3,4},
  Pamela Wyatt\altaffilmark{1}
}
\email{diana@phas.ubc.ca}
\altaffiltext{1}{NASA Exoplanet Science Institute, Caltech, MS 100-22,
  770 South Wilson Avenue, Pasadena, CA 91125}
\altaffiltext{2}{Department of Physics \& Astronomy, University of
  British Columbia, Vancouver, BC V6T1Z1, Canada}
\altaffiltext{3}{Department of Astronomy and Astrophysics,
  Pennsylvania State University, 525 Davey Laboratory, University
  Park, PA 16802}
\altaffiltext{4}{Center for Exoplanets \& Habitable Worlds,
  Pennsylvania State University, 525 Davey Laboratory, University
  Park, PA 16802}
\altaffiltext{5}{Departamento de Astonom\'ıa y Astrof\'ısica,
  Pontificia Universidad Cat\'olica de Chile, Casilla 306, Santiago
   22, Chile}
\altaffiltext{6}{Institute for Astronomy, University of Hawai’i, 2680 Woodlawn Dr, Honolulu, HI 96822}

\begin{abstract}

The Transit Ephemeris Refinement and Monitoring Survey (TERMS)
conducts radial velocity and photometric monitoring of known
exoplanets in order to refine planetary orbits and predictions of possible transit times. This effort is primarily
directed towards planets not known to transit, but a small sample of
our targets consist of known transiting systems. Here we present
precision photometry for 6 WASP (Wide Angle Search for Planets) planets acquired during their transit windows. We perform a Markov Chain Monte Carlo (MCMC) analysis for each planet and combine these data with previous measurements to redetermine the period and ephemerides for these planets. These observations provide recent mid-transit times which are useful for scheduling future observations.
Our results improve the ephemerides of WASP-4b, WASP-5b and WASP-6b and reduce the uncertainties on the mid-transit time for WASP-29b. We also confirm the orbital, stellar and planetary parameters of all 6 systems. 

\end{abstract}


\section{Introduction}

Over 100 transiting planets have now been discovered and confirmed,
and the Kepler mission has contributed over 1200 additional
candidates \citep{Bor11}. Transit observations provide a wealth of
information about a planet's physical properties. The majority of known
transiting exoplanets have arisen from ground-based surveys such as
SuperWASP \citep{Poll06}, HATNet \citep{Bak04}, TrES \citep{ODon06} and
XO \citep{McC05}.

The majority of these planets are hot jupiters and are
anything but standardized compared to each other, with respect to their density and orbital
parameters. In particular several inflated hot Jupiters have been
discovered (such as HD 209458 b, first detected by \cite{Cha00}), with large radii relative to their masses. One explanation
for this effect is tidal heating, which has been shown to inflate the
radii of close-in planets even when the orbital eccentricity is very
small (\cite{Bod01}, \cite{Mill09}). Other proposed mechanisms include enhanced
atmospheric opacities \citep{Burr07} and ohmic dissipation \citep{Bat10}, but
none seem to be able to account for the entire sample of inflated hot jupiters.

The Transit Ephemeris Refinement and Monitoring Survey (TERMS) is a
project which aims to refine the orbital parameters of
intermediate-long period radial velocity planets and detect their
transits \citep{Kan09}. We present photometry for 6 of the WASP planets. These data were acquired in order to demonstrate the precision we can
achieve for these and other TERMS targets in general with the CTIO
1.0m telescope, and to test the data reduction pipeline. All 6 are
gas giants, with WASP-4b, WASP-5b, WASP-6b and WASP-6b orbiting G-type stars while WASP-29b
orbits a K dwarf and WASP-31b orbits a F star. The six host stars have apparent magnitudes in the range 11.3 $<$ m$_{V}<$ 12.6. Each system is interesting in its own right.
WASP-4b \citep{Wil08}, WASP-6b \citep{Gil09b} and WASP-31b \citep{And10} are low-density, inflated planets. Possible indications of transit timing variations \citep{Ago05} have been reported for WASP-5b \citep{Gil09a, Fuk11} since its discovery \citep{And08}.
WASP-19b \citep{Heb10} has the shortest period of all the known hot Jupiters, 
and is thus extremely interesting from a dynamical point of view. WASP-29b \citep{Hell10} is one of only a handful of Saturn-mass (and radius) planets.

In this paper we independently derive parameters for these planets and their
host stars, and provide updated ephemerides for each system based on the observed transits.

In section 2 we describe the photometric observations and
their reduction. The analysis of these data and the results are
presented in sections 3 and 4 respectively. We summarize and discuss
the results in section 5.

\section{Observations and Data Reduction}

The observations\footnote{ The photometry presented in this paper will be made publically available through the NASA Star and Exoplanet Database (NStED) at http://nsted.ipac.caltech.edu} were carried out at Cerro Tololo Inter-American
Observatory (CTIO), with most of the data acquired using the 1.0m
telescope and the Y4KCam CCD detector. The field-of-view of this
instrument is 20'$\times$20' and the readout time of the detector is 51 seconds. The earlier of two light curves for WASP-4 was obtained using the 0.9m telescope which has a CCD with a 13.5'$\times$13.5' field-of-view and readout time of 53 seconds.

WASP-4 was observed on the night of October 12, 2008
using a Johnson V-band filter, and the night of September 10, 2009
with a Cousins R-band filter. The remaining 5 WASP targets were observed through a
Cousins R-band filter.

The photometry for WASP-5 was obtained on the nights of August 31, 2009, and of September 8, 2010.

\begin{figure}[!t]
\begin{flushleft}
\includegraphics[scale=0.55]{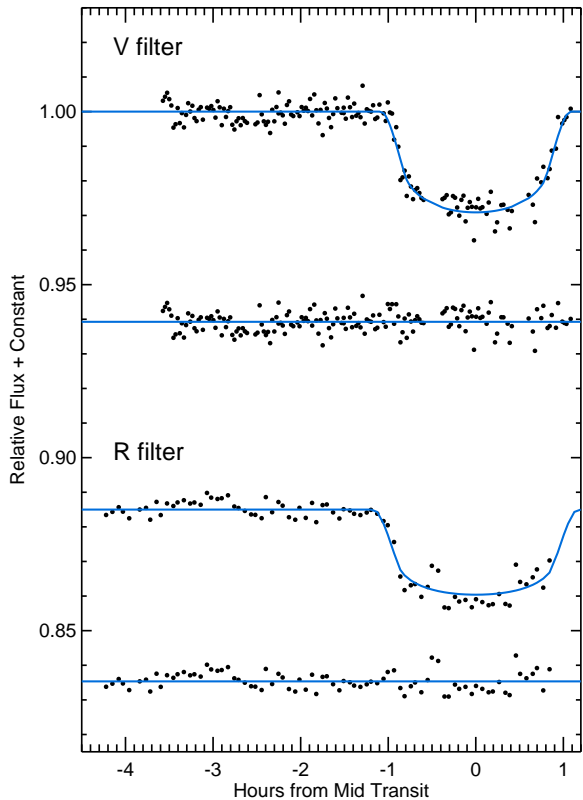}
\caption{CTIO 0.9m V-band (top) and 1.0m R-band (bottom) transit photometry of WASP-4b. The blue line is the best-fit transit light curve. The residuals are shown below each transit ({\it rms} = 2900 ppm for the V-band transit, and {\it rms} = 2600 ppm for the R-band transit). See Tables 3 and 4 for the WASP-4 system parameters. }
\end{flushleft}
\end{figure}

The photometry for WASP-6 was obtained on the night of September 6,
2010. The photometry for WASP-19 was obtained on the night of
January 18, 2011. WASP-29 was observed on the night of September 5, 2010. WASP-31 was observed on the night of January 25, 2011. 
Detailed information pertaining to the observations can be found in Table 1.

All images were bias subtracted and flat-fielded using skyflats. To increase the signal-to-noise per pixel and prevent individual pixel counts from exceeding the linear response range of the CCD, we defocussed the images to produce a relatively large point spread function, similar to that described by \cite{Sou09a}. The excellent guiding capabilities of the telescope allowed the stellar profile to remain on or near the same pixels such that accurate aperture photometry was able to be performed. Aperture radii that corresponded 
to the extent of the point spread function, and
sky annuli chosen so not to include any nearby stars were used for the
photometry. We found that the photometry is relatively robust
against the choice of these aperture radii. Between two and four
carefully selected reference stars were used in each case. Reference
stars were generally fainter than the target. Brighter stars were
often saturated (the 0.9m and the 1.0m CCDs are linear up $\sim$40000 ADU above bias) because exposure times were optimized to obtain
maximum flux for the target while maintaining the peak counts of the
PSF within the linear regime of the CCD. Relative photometry was
performed using the methods described in \cite{EH01}.

After performing aperture photometry on the transit of WASP-6b, a trend was still visible in the light curve. The trend correlates with airmass and is likely due to different spectral types of the reference stars relative to the target. Since the algorithm we used to fit the light curves (see \S~\ref{sec:ana}) can only perform a linear correction for the airmass, we removed it by fitting a 2nd order polynomial to the out-of-transit data (using time as the independent parameter) and subtracting it from the entire light curve.

\begin{figure}[!t]
\begin{flushleft}
\includegraphics[scale=0.55]{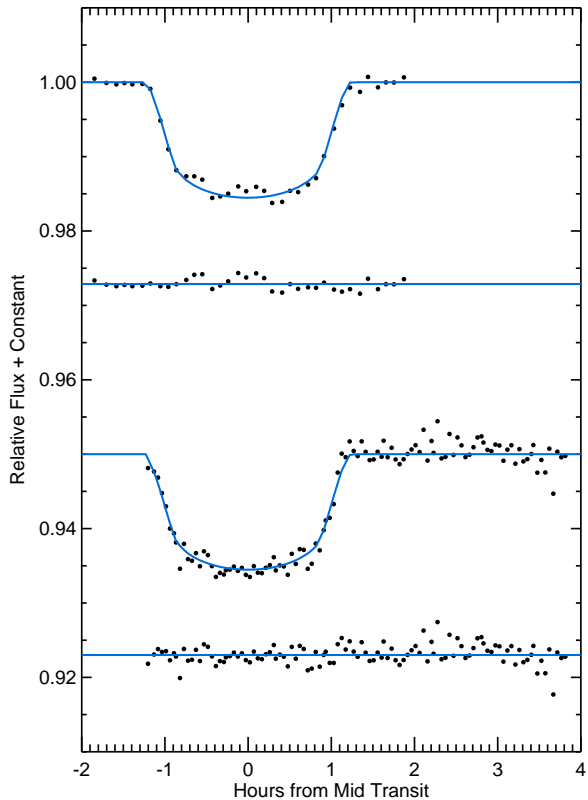}
\caption{CTIO 1.0m R-band (bottom) transit photometry of WASP-5b, observed on August 31, 2009 (top) and September 8, 2010 (bottom). The blue line is the best-fit transit light curve. The residuals are shown below each transit ({\it rms} = 700 ppm for the top transit, and {\it rms} = 1300 ppm for the bottom transit). See Table 5 for the WASP-5 system parameters.}
\end{flushleft}
\end{figure}


\section{Analysis}
\label{sec:ana}

We used TAP (Transit Analysis Package) developed by \cite{gaz11} to
obtain stellar and planetary parameters from our new light
curves. TAP accepts single or multiple light curves as input. The
light curves are then fitted using the \cite{Man02} model by means
of a Markov Chain Monte Carlo analysis. 
In the case of multiple
transits, the parameters for each transit can either evolve together
or independently. This allows, for example, to fit the transit
midpoint of each transit separately while a single value is obtained
for each of the remaining fitted parameters. We ran TAP with the eccentricity ({\it e})
and the argument of periastron ({\it $\omega$}) fixed to previously published
values because these parameters are generally more accurately determined by radial velocity measurements, and because it is difficult to constrain their values using only one or two transits for each target. 

The planet-to-star radius ratio ({\it R$_{P}$/R$_{*}$}), the scaled semi-major axis
({\it a/R$_{*}$}), the inclination ({\it i}) and the time of mid-transit ({\it T$_{0}$}) 
were allowed to float. The period was fixed to the most recent published value (as of May 2011) for each planet. We have also allowed the algorithm to fit for two additional parameters:
 uncorrelated (white) noise and correlated (red) noise. TAP uses a wavelet likelihood function to more robustly estimate 
parameter uncertainties, particularly in cases where the light curve is affected
by correlated noise \citep{Car09}. 
For each WASP light curve, we ran 10 chains of 10$^{6}$ steps each, and removed the initial 10$\%$ of each chain to account for
burn-in.

We have analysed all of our data with fixed limb darkening
coefficients (assuming a quadratic limb darkening law), using values
interpolated from the tables of \cite{Cla00} appropriate for each
star. We have also repeated the analysis with {\it u$_{1}$} and {\it
  u$_{2}$} as free parameters. Since those two parameters are
correlated, the uncorrelated linear combinations {\it 2u$_{1} +$ u$_{2}$}
and {\it u$_{1} -$ 2u$_{2}$} were fitted instead \citep{Hol06}.
For all six systems, the values of the parameters obtained when the limb darkening 
coefficients were allowed to float are only slightly different than those obtained by fixing 
the limb darkening coefficients, and they agree within uncertainties. In all cases, the fitted values of {\it u$_{1}$} and 
{\it u$_{2}$} had large uncertainties ($\sim \pm$ 0.3 for {\it u$_{1}$} and $\sim \pm$ 0.4 for {\it u$_{2}$}), indicating that our photometry does not improve upon the known limb darkening properties of these stars. These values were consistent with the \cite{Cla00} values within uncertainties.

Hence the parameter values we report have been obtained by fixing the values of the limb darkening coefficients, except for one of the WASP-4b analyses.

While we acquired one transit for each of WASP-6b, WASP-19b, WASP-29b and WASP-31b, we observed and analysed two transits for each of WASP-4b and WASP-5b. For WASP-4b, we have analysed the light curves separately since they were taken using different filters. However, we have also performed a combined analysis of the two data sets in order to obtain more accurate and precise estimates of the stellar and planetary parameters. In this case we allowed the limb darkening coefficients to float, since fixing them is inappropriate due to the different filters. For completeness, we report the results of all three analyses of the WASP-4b light curves (Tables 3 and 4). In the case of WASP-5b, we have analysed both light curves together and derived a single value for each transit parameter except the mid-transit time (for which we obtained one value per transit).

Our transit light curves for the six planets are shown in Figures 1 to 6, respectively.


\begin{figure}[!t]
\begin{flushleft}
\includegraphics[scale=0.55]{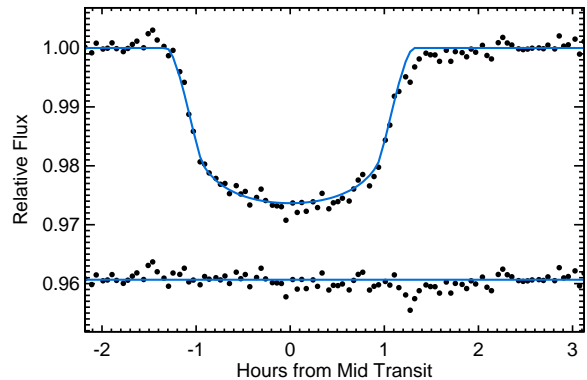}
\caption{CTIO 1.0m R-band transit photometry of WASP-6b, after airmass correction (see text for details). The blue line is the best-fit transit light curve. The residuals are shown at the bottom of the figure ({\it rms} = 1300 ppm). See Table 6 for the WASP-6 system parameters.}
\end{flushleft}
\end{figure}


\section{Results}

\subsection{New mid-transit times and ephemerides}

Since their discovery, new transit times have been published for both of WASP-4b and WASP-5b (\cite{Gil09a}, \cite{Sou09a}, \cite{Sou09b}, \cite{Tri10}, \cite{Fuk11}). This wealth of data naturally invites transit timing variation analyses. However, we believe that for such an analysis to be consistent and as accurate as possible, all available transit photometry for a given system should be fitted using the same algorithm and the results analysed using the same method. This is beyond the scope of this paper, but we encourage such studies which are now becoming increasingly possible with the advent of online exoplanet databases such as the NASA Star and Exoplanet Database \citep{Bra09}.

For the purpose of this paper, we combine the mid-transit times from our analysis with all previously published times to maximise the time span of observations, and use them to determine a new ephemeris for each system. For each of WASP-4b and WASP-5b we have acquired and analysed two transits. The two transits were observed about one year apart for each target. Together with the fifteen published values for WASP-4b (\cite{Gil09a}, \cite{Win09}, \cite{Sou09b}, \cite{San11}) we have a total of seventeen mid-transit points. For WASP-5b we combined thirteen published values (\cite{Gil09a}, \cite{Sou09a},\cite{Fuk11}) with our two for a total of fifteen mid-transit points.

For WASP-19b, we used three mid-transit times: two previously published values (\cite{Heb10}, \cite{Hell11}) and the one corresponding to the transit we have observed. \cite{Hell11} did not report the mid-transit time corresponding to the transit they obtained on the night of 2010 February 28 because they performed a combined analysis of both that transit and the one presented in \cite{Heb10}. We fitted the data obtained by \cite{Hell11} (private communication) using TAP (as described in section 3) to obtain the mid-transit time.

For each of WASP-6b, WASP-29b and WASP-31b, we have one transit which we combined with the original published values (\cite{Gil09b},\cite{Hell10} and \cite{And10}, respectively) for a total of two mid-transit points for each system.

When necessary, these times were converted from JD$_{UTC}$ (our values) and from HJD$_{UTC}$ (the discovery paper values) to BJD$_{TDB}$ using the online calculator developed by \cite{Eas10}. We determined a new ephemeris by fitting a linear function to the mid-transit points for each system:

\begin{eqnarray}
T_{0}[n]=T_{0}[0]+nP
\end{eqnarray}

We chose to center the set of mid-transit times we used for each exoplanet so as to minimise the covariance between $T_{0}[0]$ and $P$. This means that the values of $T_{0}$ reported in Tables 3 to 9 do not necessarily correspond to any of the mid-transit times in Table 2 for a specific planet.
The covariance calculated in this manner is sufficiently small for its influence in the calculation  of the uncertainty on future mid-transit times to be negligible.


\begin{figure}[!t]
\begin{flushleft}
\includegraphics[scale=0.55]{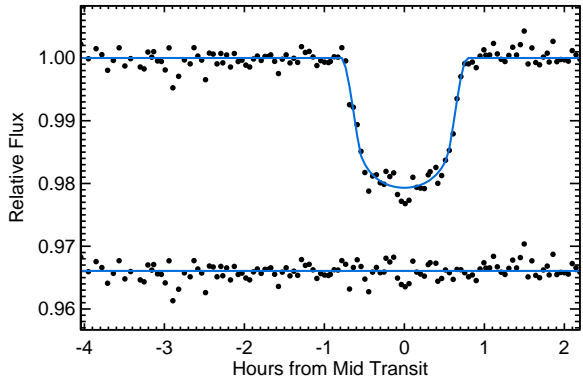}
\caption{CTIO 1.0m R-band transit photometry of WASP-19b. The blue line is the best-fit transit light curve. The residuals are shown at the bottom of the figure ({\it rms} = 1300 ppm). See Table 7 for the WASP-19 system parameters.}
\end{flushleft}
\end{figure}

\subsection{Determination of stellar and planetary properties}

For each system, the values of the orbital eccentricity, argument of periastron, velocity semi-amplitude ({\it K}), stellar mass ($M_{*}$) and planet mass ($M_{P}$) were 
taken from previous publications (see footnotes of Tables 3 to 9 for details). These values together with the light curve parameters
were used to calculate the remaining parameters in the tables. Two of those quantities, the stellar density ($\rho_{*}$) and the planet surface gravity (g$_{P}$), can be obtained solely from the photometry. For $\rho_{*}$ we used the equation

\begin{eqnarray}\label{eqn:math1}
\rho_{*} + k^{3}\rho_{\footnotesize{P}} = \frac{3\pi}{GP^{2}}\left(\frac{a}{R_{*}}\right)^{3}
\end{eqnarray}

from \cite{Sea03}, where {\it k} is a constant coefficient for each stellar sequence, which relates the stellar mass and radius, $\rho_{P}$ is the planetary density, and {\it P} is the orbital period. Since $k^{3}$ is usually small \citep{Win10}, equation \ref{eqn:math1} becomes

\begin{eqnarray}\label{eqn:math2}
\rho_{*} \approx \frac{3\pi}{GP^{2}}\left(\frac{a}{R_{*}}\right)^{3}.
\end{eqnarray}

We calculated g$_{P}$ using

\begin{eqnarray}\label{eqn:math3}
g_{P} = \frac{2\pi}{P} \frac{\sqrt{1-e^{2}}K_{*}}{(R_{P}/a)^{2}\mathrm{sin}i}
\end{eqnarray}

which is derived in \cite{Sou07}.

In Tables 3 to 9 we report our parameter estimates for the stellar and planetary properties.

\begin{figure}[!t]
\begin{flushleft}
\includegraphics[scale=0.55]{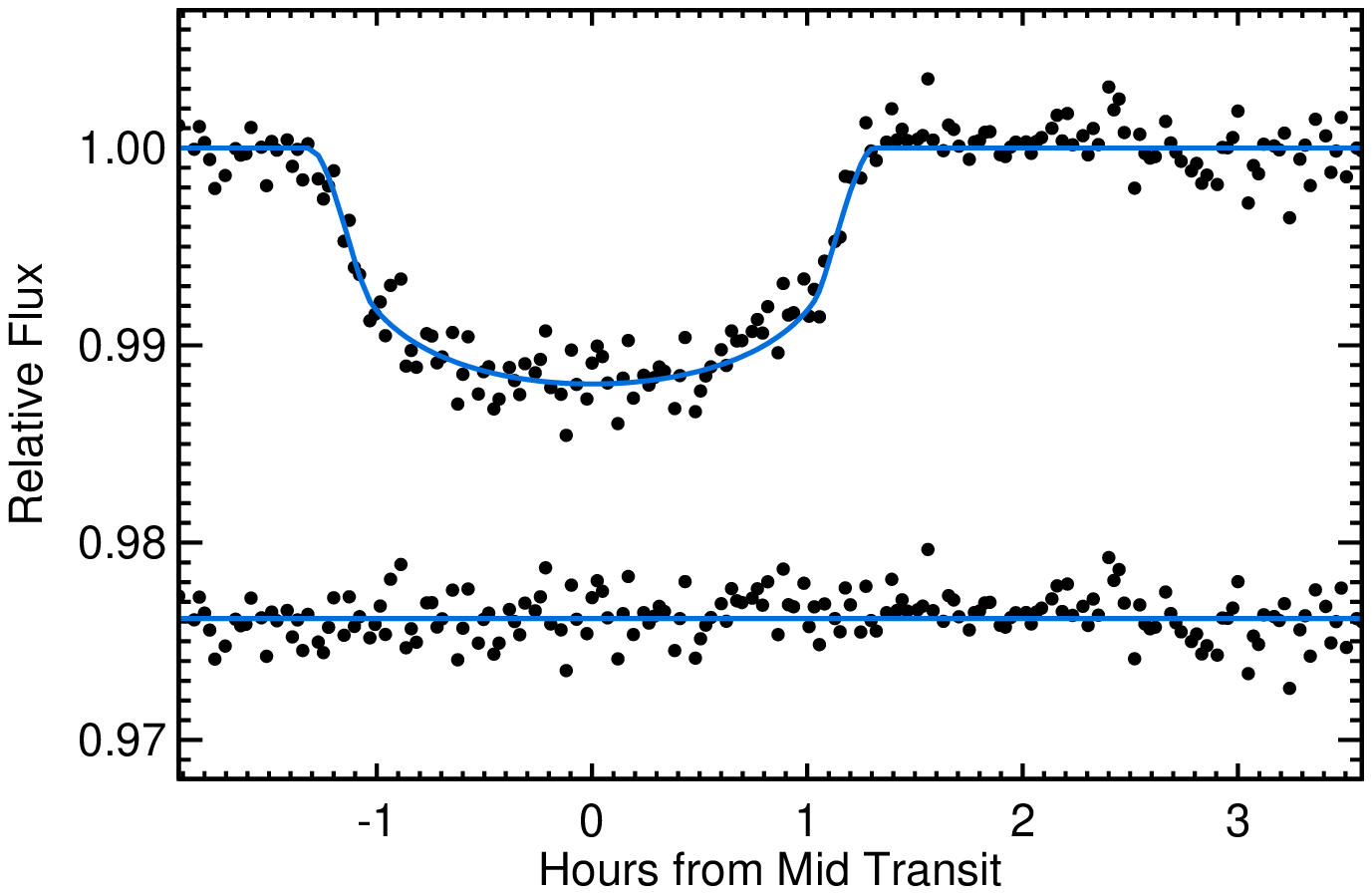}
\caption{CTIO 1.0m R-band transit photometry of WASP-29b. The blue line is the best-fit transit light curve. The residuals are shown at the bottom of the figure ({\it rms} = 1200 ppm). See Table 8 for the WASP-29 system parameters.}
\end{flushleft}
\end{figure}


\begin{figure}[!t]
\begin{flushleft}
\includegraphics[scale=0.55]{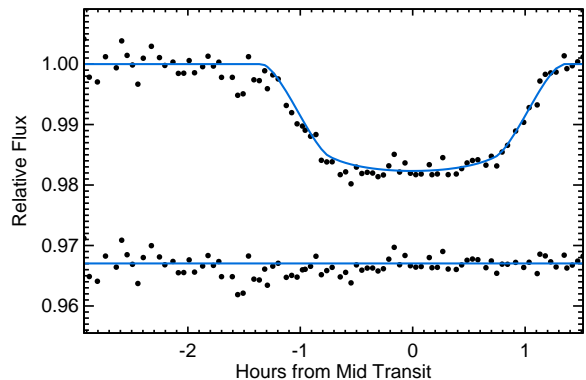}
\caption{CTIO 1.0m R-band transit photometry of WASP-31b. The blue line is the best-fit transit light curve. The residuals are shown at the bottom of the figure ({\it rms} = 1700 ppm). See Table 9 for the WASP-31 system parameters.}
\end{flushleft}
\end{figure}

\section{Discussion}

We present new photometry for six known transiting exoplanet systems: WASP-4, WASP-5, WASP-6, WASP-19, WASP-29 and WASP-31. We have reduced the data and used TAP - an interactive MCMC-based software package - to fit the transit photometry. The code robustly evaluates parameter uncertainties using a wavelet-based method for dealing with red noise in the light curves and fits for white and red noise, as well as any linear trend that may be present in the data (due to airmass for example), in addition to the transit parameters. Based on the photometry and radial velocity-derived parameters obtained from the literature, we compute and report new values for the orbital parameters, as well as the stellar and planetary properties of these six systems. Specifically, we determine the mid-transit time, the scaled semi-major axis, the planet/star area ratio and the orbital inclination solely from our new photometric measurements. The values for the orbital eccentricity, the argument of periastron, the velocity semi-amplitude, and the stellar and planetary masses are adopted from the literature (as described in the footnotes of Tables 3 to 9). The transit duration, the semi-major axis, as well as the stellar and planetary radii, densities and surface gravities were calculated from combinations of light curve parameters and parameters adopted from the literature.
 This work also constitutes the first follow-up observation paper for WASP-6b, WASP-29b and WASP-31b since publication of the discovery papers.

All our parameter values - except the period - for the WASP-4, WASP-6 and WASP-31 systems agree with those published by \cite{Tri10}, \cite{Gil09b} and \cite{And10} respectively, within the $1\sigma$ uncertainties, while the period agrees within the $2\sigma$ uncertainties in each case. 

The parameter values stemming from the combined analysis of the WASP-4 light curves (Table 4) agree better with the results of \cite{Tri10} and are more precise than the parameter values based on the individual V-band and R-band light curves (Table 3). 

All of our results for the WASP-29 system agree with those reported by \cite{Hell10} within the $1\sigma$ uncertainties, and our uncertainties on the mid-transit time are smaller. The estimates of the planetary radius are of comparable precision, confirming its Saturn-like size. For the WASP-19 system, our values for all parameters agree with those reported by \cite{Hell11} within the $2\sigma$ uncertainties. Our value for the stellar radius of WASP-5 agrees with the value reported by \cite{Fuk11} (hereafter F10) within the $2\sigma$ uncertainties, which in turn causes our value for the planet radius to be about 23 $\%$ larger than that reported in the same paper. The remaining stellar and planetary parameters agree within the $2\sigma$ uncertainties. F10 analysed seven new transits together with all previously published transit photometry for this system, and detected a deviation from a linear ephemeris which, if real, corresponds to transit timing variations (TTVs) with an amplitude of up to 50 seconds. They also find a period which does not agree with previously published values within error bars, which can be considered as a possible indication of TTVs as well. Our period value for WASP-5b is consistent with that determined by F10 within the $1\sigma$ uncertainties, but, like theirs, differs from those obtained by \cite{Sou09a} and \cite{Tri10} within error bars ($1\sigma$ uncertainties). 

We improve the ephemerides of WASP-4b, WASP-5b and WASP-6b and obtain a better constrained mid-transit time for WASP-29b. Our ephemeris for WASP-19b is of similar precision to that reported by \cite{Hell11}. Our estimates of the mid-transit times for WASP-29b and WASP-31b and those obtained by \cite{Hell10} and \cite{And10} respectively, are of comparable precision.

Having a long time span between our observations and the previous published mid-transit times, we update and refine the ephemerides for the six WASP systems we observed and analysed. Our results are thus useful for planning future observations. In addition, they are consistent with previously published parameter values and as such serve to confirm the properties of these systems.


\section*{Acknowledgements}


The authors would like to thank Andr\'{e}s Jord\'{a}n for providing support for the observations at CTIO, and the anonymous referee who provided helpful comments that improved the manuscript. We also thank C. Hellier for the ESO NTT/EFOSC light curve of WASP-19. D.D. is supported by a FQRNT scholarship and a IPAC Visiting Graduate fellowship. M.R. acknowledges support from ALMA-CONICYT projects 31090015 and 31080021. The Center for Exoplanets and Habitable Worlds is supported by the Pennsylvania State University, the Eberly College of Science, and the Pennsylvania Space Grant Consortium.


\clearpage


\begin{deluxetable*}{ccccccccc}
\tabletypesize{\scriptsize}
\tablecaption{\label {tab:obs} Log of observations}
\tablewidth{16cm}
\tablehead{
\colhead{WASP} &
\colhead{Date} &
\colhead{Telescope} &
\colhead{Filter} &
\colhead{Exp. time} &
\colhead{$\#$ of exposures} &
\colhead{$\sigma_{red}^{*}$} &
\colhead{$\sigma_{white}^{*}$} &
\colhead{RMS} \\
\colhead{exoplanet} &
\colhead{} &
\colhead{} &
\colhead{} &
\colhead{(s)} &
\colhead{} &
\colhead{(rel. flux)} &
\colhead{(rel. flux)} &
\colhead{(rel. flux)} 
}

\startdata

\\
WASP-4b & 12 Oct 2008 & CTIO 0.9m & V & 60 & 165 & 0.0110& 0.0021 &  0.0029 \\
WASP-4b & 10 Sep 2009 & CTIO 1.0m & R & 150 & 80 & 0.0139 & 0.0011 &  0.0026 \\
\\
WASP-5b & 31 Aug 2009 & CTIO 1.0m & R & 300 & 36 & 0.00262 & 0.00033 &  0.0007 \\
WASP-5b & 8 Sep 2010 & CTIO 1.0m & R & 90 & 109  & 0.0046 & 0.0011 &  0.0013 \\
\\
WASP-6b & 6 Sep 2010 & CTIO 1.0m & R & 120 & 101 & 0.00691 & 0.00056 &  0.0013 \\
\\
WASP-19b & 18 Jan 2011 & CTIO 1.0m & R & 90 & 127  & 0.0073 & 0.0010 & 0.0013 \\
WASP-29b & 5 Sep 2010 & CTIO 1.0m & R & 30 & 185 & 0.00450 & 0.00089 & 0.0012 \\
WASP-31b & 25 Jan 2011 & CTIO 1.0m & R & 90 & 87 & 0.0063 & 0.0011 &  0.0017 \\

\tablecomments{$^{*}$ $\sigma_{red}$ is the red noise parameter and $\sigma_{white}$ is the white noise parameter. The values for each were obtained using TAP. It is important to note that even though $\sigma_{white}$ is the rms of the white noise, $\sigma_{red}$ is not the rms of the red noise \citep{Car09}. }

\enddata

\end{deluxetable*}


\begin{deluxetable*}{ccccc}
\tabletypesize{\scriptsize}
\tablecaption{\label {tab:obs} Mid-transit times for all six WASP exoplanets}
\tablewidth{14cm}
\tablehead{
\colhead{WASP exoplanet} &
\colhead{Epoch} &
\colhead{Mid-transit time (BJD$_{TDB}$)}&
\colhead{Filter} &
\colhead{Reference} 
}

\startdata

\\ 
& -549 & 2453963.10863$^{+0.00074}_{-0.00081}$ & R & \cite{Gil09a}$^{*}$ \\
& -249 & 2454364.57722$^{+0.00068}_{-0.00075}$ & R & \cite{Gil09a}$^{*}$ \\
& -246 & 2454368.59244$^{+0.00022}_{-0.00019}$ & R & \cite{Gil09a} \\
& -244 & 2454371.26812$^{+0.00033}_{-0.00028}$ & I & \cite{Gil09a} \\
& -225 & 2454396.695410$^{0.000051+}_{-0.000051}$ & Z & \cite{Gil09a} \\
& 0 & 2454697.797489$^{+0.000055}_{-0.000055}$ & Z & \cite{Win09} \\
& 0 & 2454697.798219$^{+0.00010}_{-0.00010}$ & R & \cite{Sou09b} \\
& 3 & 2454701.812919$^{+0.00013}_{-0.00013}$ & R & \cite{Sou09b} \\
WASP-4b &  26 & 2454732.591919$^{0.00013+}_{-0.00013}$ & R & \cite{Sou09b} \\
& 32 & 2454740.621560$^{0.000061+}_{-0.000061}$ & R & \cite{Sou09b} \\
& 38 & 2454748.650490$^{0.000072+}_{-0.000072}$ & Z & \cite{Win09} \\
& 41 & 2454752.66591$^{+0.00071}_{-0.00071}$ & V & This work \\
& 260 & 2455045.738643$^{+0.000054}_{-0.000054}$ & Z & \cite{San11} \\
& 263 & 2455049.753274$^{+0.000066}_{-0.000066}$ & Z & \cite{San11} \\
& 266 & 2455053.767816$^{+0.000053}_{-0.000053}$ & Z & \cite{San11} \\
& 290 & 2455085.8853$^{+0.0038}_{-0.0016}$ & R & This work \\
& 301 & 2455100.605928$^{+0.000061}_{-0.000061}$ & Z & \cite{San11} \\
\\
\hline
\\
& -264 & 2453945.71962$^{+0.00091}_{-0.00093}$ & R & \cite{Gil09a}$^{*}$\\
&  -7 & 2454364.2283$^{+0.0012}_{-0.0013}$ & R & \cite{Gil09a}$^{*}$ \\
& 5 & 2454383.76751$^{0.00028+}_{-0.00028}$ & R & \cite{Gil09a} \\
& 7 & 2454387.02286$^{0.00086+}_{-0.00086}$ & I & \cite{Gil09a} \\
& 160 & 2454636.17465$^{+0.00047}_{-0.00047}$ & I & \cite{Fuk11} \\
& 204 & 2454707.82531$^{+0.00021}_{-0.00021}$ & R & \cite{Sou09a} \\
& 218 & 2454730.62252$^{+0.00022}_{-0.00022}$ & R & \cite{Sou09a} \\
WASP-5b & 244 & 2454772.96212$^{+0.00051}_{-0.00051}$ & I & \cite{Fuk11} \\
& 430 & 2455075.84936$^{+0.00055}_{-0.00056}$ & R & This work \\
& 432 & 2455079.10849$^{+0.00044}_{-0.00044}$ & I & \cite{Fuk11} \\
& 451 & 2455110.04645$^{+0.00073}_{-0.00073}$ & I & \cite{Fuk11} \\
& 459 & 2455123.07627$^{+0.00041}_{-0.00041}$ & I & \cite{Fuk11} \\
& 607 & 2455364.08262$^{+0.00057}_{-0.00057}$ & I & \cite{Fuk11} \\
& 615 & 2455377.10969$^{+0.00048}_{-0.00048}$ & I & \cite{Fuk11} \\
& 659 & 2455448.75950$^{+0.00086}_{-0.00084}$ & R & This work \\
\\
\hline
\\ WASP-6b & 0 & 2454596.43342$^{+0.00015}_{-0.00010}$ & - & \cite{Gil09b}$^{*}$ \\
& 253  & 2455446.76621$^{+0.00058}_{-0.00057}$ & R & This work \\
\\
\hline
& 0 & 2454775.33796$^{+0.00010}_{-0.00020}$ & - & \cite{Heb10}$^{*}$ \\
WASP-19b & 609 & 2455255.74105$^{+0.00014}_{-0.00015}$ & R & \cite{Hell11} (private communication) \\
& 1021 & 2455580.74124$^{+0.00057}_{-0.00059}$ & R & This work \\
\\
\hline
\\ WASP-29b & 0 & 2455320.2348$^{+0.0040}_{-0.0040}$ & - & \cite{Hell10}$^{*}$ \\
& 32  & 2455445.76247$^{+0.00073}_{-0.00070}$ & R & This work \\
\\
\hline
\\WASP-31b & 0 & 2455189.2836$^{+0.0003}_{-0.0003}$ & - & \cite{And10}$^{*}$ \\
& 117  & 2455587.7719 $^{+0.0014}_{-0.0013}$  & R & This work \\

\tablecomments{$^{*}$ The mid-transit times reported in the discovery papers for these targets were computed based on several transits.}

\enddata

\end{deluxetable*}


\begin{deluxetable*}{cccc}
\tabletypesize{\scriptsize}
\tablecaption{\label {tab:obs} System parameters for WASP-4 (V and R light curves analysed separately)}
\tablewidth{14cm}
\tablehead{
\colhead{Parameter} &
\colhead{Symbol} &
\colhead{This work (V filter)}  &
\colhead{This work (R filter)} 
}

\startdata
Mid-transit time (BJD$_{TDB}$) & {\it T$_0$} & 2454823.591767$^{+0.000019}_{-0.000019}$ & 2454823.591767$^{+0.000019}_{-0.000019}$  \\
Orbital period (days) & {\it P} & 1.33823326$^{+0.00000011}_{-0.00000011}$ & 1.33823326$^{+0.00000011}_{-0.00000011}$  \\
Scaled semimajor axis & {\it a/R$_{*}$} & 5.45$^{+0.15}_{-0.24}$ & 4.96$^{+0.38}_{-0.62}$ \\
Planet/star area ratio & ({\it R$_p$/R$_{*}$})$^{2}$ & 0.0233$^{+0.0017}_{-0.0015}$ & 0.0205$^{+0.0030}_{-0.0030}$  \\
Orbital inclination (deg) & {\it i} & 88.2$^{+1.2}_{-1.9}$ & 87.2$^{+2.0}_{-2.9}$  \\
Transit duration$^{1}$ (hours) & $t_{T}$ & 2.154$^{+0.083}_{-0.114}$ & 2.32$^{+0.19}_{-0.30}$ \\
Semimajor axis (AU) & {\it a} & 0.02320$^{+0.00042}_{-0.00042}$ & 0.02320$^{+0.00042}_{-0.00042}$  \\
Orbital eccentricity & {\it e} & 0 (adopted) & 0 (adopted)  \\
Argument of periastron (deg) & {\it $\omega$} & - & - \\
Velocity semi-amplitude (m/s) & {\it K} & 242.1$^{+2.8}_{-3.1}$ (adopted) & 242.1$^{+2.8}_{-3.1}$ (adopted)  \\ 
Stellar mass (M$_{\odot}$) & {\it M$_{*}$} & 0.93$^{+0.05}_{-0.05}$ (adopted) & 0.93$^{+0.05}_{-0.05}$ (adopted)  \\
Stellar radius (R$_{\odot}$) & {\it R$_{*}$} & 0.915$^{+0.030}_{-0.043}$ & 1.005$^{+0.079}_{-0.127}$  \\
Stellar density ($g/cm^{3}$) & {\it $\rho_{*}$} & 1.71$^{+0.19}_{-0.26}$ & 1.29$^{+0.31}_{-0.49}$  \\
Stellar surface gravity (cgs) & log {\it g$_{*}$} & 4.484$^{+0.037}_{-0.047}$ & 4.402$^{+0.072}_{-0.112}$  \\
Planet mass (M$_{Jup}$) & {\it M$_{P}$} & 1.250$^{+0.050}_{-0.051}$ (adopted) & 1.250$^{+0.050}_{-0.051}$ (adopted)  \\
Planet radius (R$_{Jup}$) & {\it R$_{P}$} & 1.389$^{+0.068}_{-0.080}$ & 1.43$^{+0.15}_{-0.21}$  \\
Planet density ($g/cm^{3}$) & {\it $\rho_{P}$} & 0.619$^{+0.095}_{-0.110}$ & 0.56$^{+0.18}_{-0.25}$  \\
Planet surface gravity (cgs) & log {\it g$_{P}$} & 3.225$^{+0.046}_{-0.050}$ & 3.199$^{+0.093}_{-0.125}$

\tablecomments{$^{1}$The transit duration is from 1st contact to 4th contact. Uncertainties correspond to the 68.3$\%$ ($1\sigma$) confidence limits.. The orbital eccentricity, argument of periastron, velocity semiamplitude, stellar mass and planet mass for WASP-4 were taken from \cite{Tri10}.}

\enddata

\end{deluxetable*}


\begin{deluxetable*}{cccc}
\tabletypesize{\scriptsize}
\tablecaption{\label {tab:obs} System parameters for WASP-4 (V and R light curves analysed together)}
\tablewidth{14cm}
\tablehead{
\colhead{Parameter} &
\colhead{Symbol} &
\colhead{This work (combined)} &
\colhead{\cite{Tri10}} 
}

\startdata
Mid-transit time (BJD$_{TDB}$) & {\it T$_0$} & 2454823.591767$^{+0.000019}_{-0.000019}$ & 2454387.327787$^{+0.000040}_{-0.000039}$ \\
Orbital period (days) & {\it P} & 1.33823326$^{+0.00000011}_{-0.00000011}$  & 1.3382299$^{+0.0000023}_{-0.0000021}$ \\
Scaled semimajor axis & {\it a/R$_{*}$} & 5.53$^{+0.16}_{-0.21}$ & 5.5313$^{+0.011}_{-0.012}$ \\
Planet/star area ratio & ({\it R$_p$/R$_{*}$})$^{2}$ & 0.0230485$^{+0.0016}_{-0.0017}$. & 0.023333$^{+0.000043}_{-0.000073}$ \\
Orbital inclination (deg) & {\it i} & 88.5$^{+1.0}_{-1.6}$ & 89.47$^{+0.51}_{-0.24}$ \\
Transit duration$^{1}$ (hours) & $t_{T}$ & 2.128$^{+0.080}_{-0.096}$ & 2.1283$^{+0.0019}_{-0.0003}$ \\
Semimajor axis (AU) & {\it a} & 0.02320$^{+0.00042}_{-0.00042}$ & 0.02320$^{+0.00044}_{-0.00045}$ \\
Orbital eccentricity & {\it e} & 0 (adopted) & $<$ 0.0182 \\
Argument of periastron (deg) & {\it $\omega$} & - & - \\
Velocity semi-amplitude (m/s) & {\it K} & 242.1$^{+2.8}_{-3.1}$ (adopted) & 242.1$^{+2.8}_{-3.1}$ \\ 
Stellar mass (M$_{\odot}$) & {\it M$_{*}$} & 0.93$^{+0.05}_{-0.05}$ (adopted)  & 0.93$^{+0.05}_{-0.05}$  \\
Stellar radius (R$_{\odot}$) & {\it R$_{*}$} & 0.902$^{+0.031}_{-0.038}$ & 0.903$^{+0.016}_{-0.019}$ \\
Stellar density ($g/cm^{3}$) & {\it $\rho_{*}$} & 1.79${+0.21}_{-0.24}$ & 1.786$^{+0.012}_{-0.011}$ \\
Stellar surface gravity (cgs) & log {\it g$_{*}$} & 4.496$^{+0.038}_{-0.043}$ & 4.5$^{+0.2}_{-0.2}$ \\
Planet mass (M$_{Jup}$) & {\it M$_{P}$} & 1.250$^{+0.050}_{-0.051}$ (adopted) & 1.250$^{+0.050}_{-0.051}$ \\
Planet radius (R$_{Jup}$) & {\it R$_{P}$} & 1.363$^{+0.066}_{-0.076}$ & 1.341$^{+0.023}_{-0.029}$ \\
Planet density ($g/cm^{3}$) & {\it $\rho_{P}$} & 0.655$^{+0.098}_{-0.112}$ & - \\
Planet surface gravity (cgs) & log {\it g$_{P}$} & 3.242$^{+0.045}_{-0.048}$ & -

\tablecomments{$^{1}$The transit duration is from 1st contact to 4th contact. Uncertainties correspond to the 68.3$\%$ ($1\sigma$) confidence limits.. The orbital eccentricity, argument of periastron, velocity semiamplitude, stellar mass and planet mass for WASP-4 were taken from \cite{Tri10}.}

\enddata

\end{deluxetable*}

\begin{deluxetable*}{cccc}
\tabletypesize{\scriptsize}
\tablecaption{\label {tab:obs} System parameters for WASP-5}
\tablewidth{15cm}
\tablehead{
\colhead{Parameter} &
\colhead{Symbol} &
\colhead{This work} &
\colhead{\cite{Fuk11}} 
}

\startdata

Mid-transit time (BJD$_{TDB}$) & {\it T$_0$} & 2454782.73290$^{+0.00010}_{-0.00010}$ & 2454375.62510$^{+0.00019}_{-0.00019}$ \\
Orbital period (days) & {\it P} & 1.62843064$^{+0.00000057}_{-0.00000057}$ & 1.62843142$^{+0.00000064}_{-0.00000064}$ \\
Scaled semimajor axis & {\it a/R$_{*}$} & 4.78$^{+0.30}_{-0.23}$ & 5.37$^{+0.15}_{-0.15}$ \\
Planet/star area ratio & ({\it R$_p$/R$_{*}$})$^{2}$ & 0.01392$^{+0.00066}_{-0.00070}$ & 0.01228$^{+0.00024}_{-0.00024}$ \\
Orbital inclination (deg) & {\it i} & 83.0$^{+1.4}_{-1.1}$ & 85.58$^{+0.81}_{-0.76}$ \\
Transit duration$^{1}$ (hours) & $t_{T}$ & 2.49$^{+0.29}_{-0.23}$ & - \\
Semimajor axis (AU) & {\it a} & 0.02709$^{+0.00057}_{-0.00058}$ & 0.02702$^{+0.00059}_{-0.00059}$ \\
Orbital eccentricity & {\it e} & 0 (adopted) & - \\
Argument of periastron (deg) & {\it $\omega$} & - & - \\
Velocity semi-amplitude (m/s) & {\it K} & 268.7$^{+1.8}_{-1.9}$ (adopted) & - \\ 
Stellar mass (M$_{\odot}$) & {\it M$_{*}$} & 1.000$^{+0.063}_{-0.064}$ (adopted) & - \\
Stellar radius (R$_{\odot}$) & {\it R$_{*}$} & 1.218$^{+0.081}_{-0.064}$ & 1.082$^{+0.038}_{-0.038}$  \\
Stellar density ($g/cm^{3}$) & {\it $\rho_{*}$} & 0.78$^{+0.16}_{-0.13}$ & 1.11$^{+0.14}_{-0.14}$ \\
Stellar surface gravity (cgs) & log {\it g$_{*}$} & 4.267$^{+0.064}_{-0.053}$ &  \\
Planet mass (M$_{Jup}$) & {\it M$_{P}$} & 1.555$^{+0.066}_{-0.072}$ (adopted) & 1.568$^{+0.071}_{-0.071}$  \\
Planet radius (R$_{Jup}$) & {\it R$_{P}$} & 1.430$^{+0.100}_{-0.083}$ & 1.167$^{+0.043}_{-0.043}$ \\
Planet density ($g/cm^{3}$) & {\it $\rho_{P}$} & 0.71$^{+0.15}_{-0.13}$ & 1.22$^{+0.15}_{-0.15}$ \\
Planet surface gravity (cgs) & log {\it g$_{P}$} & 3.298$^{+0.064}_{-0.051}$ & -

\tablecomments{$^{1}$The transit duration is from 1st contact to 4th contact. Uncertainties correspond to the 68.3$\%$ ($1\sigma$) confidence limits. The orbital eccentricity, argument of periastron, velocity semiamplitude, stellar mass and planet mass for WASP-5 were taken from \cite{Tri10}.}

\enddata

\end{deluxetable*}


\begin{deluxetable*}{cccc}
\tabletypesize{\scriptsize}
\tablecaption{\label {tab:obs} System parameters for WASP-6}
\tablewidth{15cm}
\tablehead{
\colhead{Parameter} &
\colhead{Symbol} &
\colhead{This work} &
\colhead{\cite{Gil09b}} 
}

\startdata

Mid-transit time (BJD$_{TDB}$) & {\it T$_0$} & 2454633.40441$^{+0.00012}_{-0.00012}$ & 2454596.43341$^{+0.00015}_{-0.00010}$ \\
Orbital period (days) & {\it P} & 3.3609992$^{+0.0000023}_{-0.0000023}$ & 3.3610060$^{+0.0000022}_{-0.0000035}$ \\
Scaled semimajor axis & {\it a/R$_{*}$} & 10.18$^{+0.62}_{-0.80}$ & - \\
Planet/star area ratio & ({\it R$_p$/R$_{*}$})$^{2}$ & 0.0223$^{+0.0013}_{-0.0012}$ & 0.02092$^{+0.00019}_{-0.00025}$\\
Orbital inclination (deg) & {\it i} & 87.9$^{+1.3}_{-1.1}$ & 88.47$^{+0.65}_{-0.47}$ \\
Transit duration$^{1}$ (hours) & $t_{T}$ & 2.75$^{+0.27}_{-0.30}$ & 2.606$^{+0.018}_{-0.016}$ \\
Semimajor axis (AU) & {\it a} & 0.04208$^{+0.00080}_{-0.00128}$ & 0.0421$^{+0.0008}_{-0.0013}$ \\
Orbital eccentricity & {\it e} & 0.054$^{+0.018}_{-0.017}$ (adopted) & 0.054$^{+0.018}_{-0.017}$ \\
Argument of periastron (deg) & {\it $\omega$} & 97.4$^{+6.9}_{-13.2}$ (adopted) & 97.4$^{+6.9}_{-13.2}$ \\
Velocity semi-amplitude (m/s) & {\it K} & 74.3$^{+1.7}_{-1.4}$ (adopted) & 74.3$^{+1.7}_{-1.4}$  \\ 
Stellar mass (M$_{\odot}$) & {\it M$_{*}$} & 0.880$^{+0.050}_{-0.080}$ (adopted) & 0.880$^{+0.050}_{-0.080}$ \\
Stellar radius (R$_{\odot}$) & {\it R$_{*}$} & 0.889$^{+0.057}_{-0.075}$ & 0.870$^{+0.025}_{-0.036}$ \\
Stellar density ($g/cm^{3}$) & {\it $\rho_{*}$} & 1.77$^{+0.35}_{-0.47}$ & 1.89$^{+0.16}_{-0.14}$ \\
Stellar surface gravity (cgs) & log {\it g$_{*}$} & 4.485$^{+0.061}_{-0.083}$  & 4.50$^{+0.06}_{-0.06}$  \\
Planet mass (M$_{Jup}$) & {\it M$_{P}$} & 0.503$^{+0.019}_{-0.038}$ (adopted) & 0.503$^{+0.019}_{-0.038}$  \\
Planet radius (R$_{Jup}$) & {\it R$_{P}$} & 1.321$^{+0.092}_{-0.12}$ & 1.224$^{+0.051}_{-0.052}$  \\
Planet density ($g/cm^{3}$) & {\it $\rho_{P}$} & 0.289$^{+0.062}_{-0.080}$ & 0.36$^{+0.07}_{-0.07}$\\
Planet surface gravity (cgs) & log {\it g$_{P}$} & 2.873$^{+0.064}_{-0.077}$ & 2.940$^{+0.063}_{-0.063}$

\tablecomments{$^{1}$The transit duration is from 1st contact to 4th contact. Uncertainties correspond to the 68.3$\%$ ($1\sigma$) confidence limits. The orbital eccentricity, argument of periastron, velocity semiamplitude, stellar mass and planet mass for WASP-6 were taken from \cite{Gil09b}.}

\enddata

\end{deluxetable*}


\begin{deluxetable*}{cccc}
\tabletypesize{\scriptsize}
\tablecaption{\label {tab:obs} System parameters for WASP-19}
\tablewidth{15cm}
\tablehead{
\colhead{Parameter} &
\colhead{Symbol} &
\colhead{This work} &
\colhead{\cite{Hell11}} 
}

\startdata

Mid-transit time (BJD$_{TDB}$) & {\it T$_0$} & 2455041.96557$^{+0.00010}_{-0.00010}$ & 2455168.96879$^{+0.00009}_{-0.00009}$ \\
Orbital period (days) & {\it P} & 0.78883889$^{+0.00000032}_{-0.00000032}$ & 0.7888400$^{+0.0000003}_{-0.0000003}$ \\
Scaled semimajor axis & {\it a/R$_{*}$} & 4.02$^{+0.39}_{-0.38}$ & - \\
Planet/star area ratio & ({\it R$_p$/R$_{*}$})$^{2}$ & 0.0180$^{+0.0014}_{-0.0013}$ & 0.0206$^{+0.0002}_{-0.0002}$ \\
Orbital inclination (deg) & {\it i} & 83.0$^{+3.8}_{-2.8}$ & 79.4$^{+0.4}_{-0.4}$ \\
Transit duration$^{1}$ (hours) & $t_{T}$ & 1.55$^{+0.26}_{-0.23}$ & 1.572$^{+0.007}_{-0.007}$ \\
Semimajor axis (AU) & {\it a} & 0.01654$^{+0.00011}_{-0.00011}$ & 0.01655$^{+0.00013}_{-0.00013}$ \\
Orbital eccentricity & {\it e} & 0 (adopted) & 0.0046$^{+0.0044}_{-0.0028}$ \\
Argument of periastron (deg) & {\it $\omega$} & - & 3$^{+70}_{-70}$ \\
Velocity semi-amplitude (m/s) & {\it K} & 257$^{+3}_{-3}$ (adopted) & 257$^{+3}_{-3}$ \\ 
Stellar mass (M$_{\odot}$) & {\it M$_{*}$} & 0.97$^{+0.02}_{-0.02}$ (adopted) & 0.97$^{+0.02}_{-0.02}$ \\
Stellar radius (R$_{\odot}$) & {\it R$_{*}$} & 0.885$^{+0.086}_{-0.084}$ & 0.99$^{+0.02}_{-0.02}$ \\
Stellar density ($g/cm^{3}$) & {\it $\rho_{*}$} & 1.98$^{+0.59}_{-0.58}$ & 1.400$^{+0.066}_{-0.059}$ \\
Stellar surface gravity (cgs) & log {\it g$_{*}$} & 4.531$^{+0.085}_{-0.083}$ & 4.432$^{+0.013}_{-0.013}$ \\
Planet mass (M$_{Jup}$) & {\it M$_{P}$} & 1.168$^{+0.023}_{-0.023}$ (adopted) & 1.168$^{+0.023}_{-0.023}$ \\
Planet radius (R$_{Jup}$) & {\it R$_{P}$} & 1.18$^{+0.12}_{-0.12}$ & 1.386$^{+0.032}_{-0.032}$ \\
Planet density ($g/cm^{3}$) & {\it $\rho_{P}$} & 0.94$^{+0.29}_{-0.28}$ & 0.581$^{+0.037}_{-0.037}$ \\
Planet surface gravity (cgs) & log {\it g$_{P}$} & 3.330$^{+0.091}_{-0.088}$ & 3.143$^{+0.018}_{-0.018}$

\tablecomments{$^{1}$The transit duration is from 1st contact to 4th contact. Uncertainties correspond to the 68.3$\%$ ($1\sigma$) confidence limits. The orbital eccentricity, velocity semiamplitude, stellar mass and planet mass for WASP-19 were taken from \cite{Hell11}.}

\enddata

\end{deluxetable*}


\begin{deluxetable*}{cccc}
\tabletypesize{\scriptsize}
\tablecaption{\label {tab:obs} System parameters for WASP-29}
\tablewidth{15cm}
\tablehead{
\colhead{Parameter} &
\colhead{Symbol} &
\colhead{This work} &
\colhead{\cite{Hell10}} 
}

\startdata

Mid-transit time (BJD$_{TDB}$) & {\it T$_0$} & 2455441.83973$^{+0.00070}_{-0.00070}$ & 2455320.2348$^{+0.0040}_{-0.0040}$  \\
Orbital period (days) & {\it P} & 3.92274$^{+0.00013}_{-0.00013}$ & 3.922727$^{+0.000004}_{-0.000004}$ \\
Scaled semimajor axis & {\it a/R$_{*}$} & 11.98$^{+0.69}_{-1.25}$ & - \\
Planet/star area ratio & ({\it R$_p$/R$_{*}$})$^{2}$ & 0.00977$^{+0.00079}_{-0.00061}$  & 0.0102$^{+0.0004}_{-0.0004}$ \\
Inclination (deg) & {\it i} & 88.3$^{+1.1}_{-1.2}$ & 88.8$^{+0.7}_{-0.7}$ \\
Transit duration$^{1}$ (hours) & $t_{T}$ & 2.60$^{+0.25}_{-0.36}$ & 2.659$^{+0.036}_{-0.036}$ \\
Semimajor axis (AU) & {\it a} & 0.04566$^{+0.00061}_{-0.00061}$ & 0.0457$^{+0.0006}_{-0.0006}$ \\
Eccentricity & {\it e} & 0 (adopted) & 0.03$^{+0.05}_{-0.03}$ \\
Argument of periastron (deg) & {\it $\omega$} & - & - \\
Velocity semi-amplitude (m/s) & {\it K} & 35.6$^{+2.7}_{-2.7}$ (adopted) & 35.6$^{+2.7}_{-2.7}$ \\ 
Stellar mass (M$_{\odot}$) & {\it M$_{*}$} & 0.825$^{+0.033}_{-0.033}$ (adopted) & 0.825$^{+0.033}_{-0.033}$ \\
Stellar radius (R$_{\odot}$) & {\it R$_{*}$} & 0.820$^{+0.048}_{-0.086}$ & 0.808$^{+0.044}_{-0.044}$ \\
Stellar density ($g/cm^{3}$) & {\it $\rho_{*}$} & 2.11$^{+0.38}_{-0.67}$ & 2.20$^{+0.28}_{-0.32}$ \\
Stellar surface gravity (cgs) & log {\it g$_{*}$} & 4.527$^{+0.054}_{-0.093}$ & 4.54$^{+0.04}_{-0.04}$ \\
Planet mass (M$_{Jup}$) & {\it M$_{P}$} & 0.244$^{+0.020}_{-0.020}$ (adopted) & 0.244$^{+0.020}_{-0.020}$ \\
Planet radius (R$_{Jup}$) & {\it R$_{P}$} & 0.806$^{+0.058}_{-0.089}$ & 0.792$^{+0.056}_{-0.035}$ \\
Planet density ($g/cm^{3}$) & {\it $\rho_{P}$} & 0.61$^{+0.14}_{-0.21}$ & 0.65$^{+0.11}_{-0.11}$ \\
Planet surface gravity (cgs) & log {\it g$_{P}$} & 2.987$^{+0.071}_{-0.101}$ & 2.95$^{+0.05}_{-0.05}$

\tablecomments{$^{1}$The transit duration is from 1st contact to 4th contact. Uncertainties correspond to the 68.3$\%$ ($1\sigma$) confidence limits. The orbital eccentricity, argument of periastron, velocity semiamplitude, stellar mass and planet mass for WASP-29 were taken from \cite{Hell10}.}

\enddata

\end{deluxetable*}


\begin{deluxetable*}{cccc}
\tabletypesize{\scriptsize}
\tablecaption{\label {tab:obs} System parameters for WASP-31}
\tablewidth{15cm}
\tablehead{
\colhead{Parameter} &
\colhead{Symbol} &
\colhead{This work} &
\colhead{\cite{And10}} 
}

\startdata

Mid-transit time (BJD$_{TDB}$) & {\it T$_0$} & 2455209.71890$^{+0.00029}_{-0.00029}$ & 2455189.2836$^{+0.0003}_{-0.0003}$  \\
Orbital period (days) & {\it P} & 3.40588291$^{+0.000012}_{-0.000012}$ & 3.405909$^{+0.000005}_{-0.000005}$ \\
Scaled semimajor axis & {\it a/R$_{*}$} & 8.52$^{+1.04}_{-0.81}$ & - \\
Planet/star area ratio & ({\it R$_p$/R$_{*}$})$^{2}$ & 0.0171$^{+0.0016}_{-0.0015}$ & 0.01622$^{+0.00032}_{-0.00032}$ \\
Inclination (deg) & {\it i} & 85.17$^{+1.09}_{-0.93}$ & 84.54$^{+0.27}_{-0.27}$ \\
Transit duration$^{1}$ (hours) & $t_{T}$ & 2.67$^{+0.64}_{-0.52}$ & 2.657$^{+0.034}_{-0.034}$\\
Semimajor axis (AU) & {\it a} & 0.04657$^{+0.00035}_{-0.00035}$ & 0.04657$^{+0.00034}_{-0.00034}$ \\
Eccentricity & {\it e} & 0 (adopted) & 0 (adopted) \\
Argument of periastron (deg) & {\it $\omega$} & - & - \\
Velocity semi-amplitude (m/s) & {\it K} & 58.2$^{+3.5}_{-3.5}$ (adopted) & 58.2$^{+3.5}_{-3.5}$ \\ 
Stellar mass (M$_{\odot}$) & {\it M$_{*}$} & 1.161$^{+0.026}_{-0.026}$ (adopted) & 1.161$^{+0.026}_{-0.026}$ \\
Stellar radius (R$_{\odot}$) & {\it R$_{*}$} & 1.18$^{+0.14}_{-0.11}$ & 1.241$^{+0.039}_{-0.039}$ \\
Stellar density ($g/cm^{3}$) & {\it $\rho_{*}$} & 1.01$^{+0.37}_{-0.29}$ & 0.857$^{+0.073}_{-0.073}$ \\
Stellar surface gravity (cgs) & log {\it g$_{*}$} & 4.363$^{+0.107}_{-0.083}$ & 4.316$^{+0.024}_{-0.024}$ \\
Planet mass (M$_{Jup}$) & {\it M$_{P}$} & 0.478$^{+0.030}_{-0.030}$ (adopted) & 0.478$^{+0.030}_{-0.030}$ \\
Planet radius (R$_{Jup}$) & {\it R$_{P}$} & 1.53$^{+0.20}_{-0.16}$ & 1.537$^{+0.060}_{-0.060}$ \\
Planet density ($g/cm^{3}$) & {\it $\rho_{P}$} & 0.177$^{+0.070}_{-0.057}$ & 0.175$^{+0.023}_{-0.023}$ \\
Planet surface gravity (cgs) & log {\it g$_{P}$} & 2.724$^{+0.117}_{-0.095}$ & 2.665$^{+0.042}_{-0.042}$

\tablecomments{$^{1}$The transit duration is from 1st contact to 4th contact. Uncertainties correspond to the 68.3$\%$ ($1\sigma$) confidence limits. The orbital eccentricity, argument of periastron, velocity semiamplitude, stellar mass and planet mass for WASP-31 were taken from \cite{And10}.}

\enddata

\end{deluxetable*}

\end{document}